\documentclass[a4paper,12pt]{article}
\usepackage[cp1251]{inputenc}
\usepackage{amsmath}
\usepackage{amsthm}
\usepackage{amssymb}
\usepackage{cite}
\usepackage{onspec}

\theoremstyle{definition}
\newtheorem*{definition}{Definition}

\theoremstyle{theorem}

\DeclareMathOperator{\spec}{spec}

\begin{document}

\renewcommand{\labelenumi}{(\theenumi)}
\renewcommand{\theenumi}{\roman{enumi}}

\title{\Large\textbf{On Standards and Specifications in Quantum
Cryptography}}

\author{\normalsize{A. S.
Trushechkin$^{1,2}$ and I. V.
Volovich$^1$}}\date{\normalsize\textit{$^1$Steklov Mathematical
Institute of the Russian Academy of Sciences, Moscow\\
$^2$Moscow Engineering Physics Institute, Russia\\
e-mail: trushechkin@mail.ru, volovich@mi.ras.ru}} \maketitle

\begin{abstract}
Quantum cryptography is going to find practically useful
applications. Recently some first quantum cryptographic solutions
became available on the market. For clients it is important to be
able to compare the quality and properties of the proposed
products. To this end one needs to elaborate on specifications and
standards of solutions in quantum cryptography. We propose and
discuss a list of characteristics for the specification, which
includes numerical evaluations of the security of solution and can
be considered as a standard for quantum key distribution
solutions. The list is based on the average time of key generation
depending on some parameters. In the simplest case for the user
the list includes three characteristics: the security degree
$\varepsilon$, the length of keys $m$ and the key refresh rate
$R$.
\end{abstract}

\tableofcontents

\section{Introduction}

In 1984 Bennett and Brassard \cite{BennettBrassard84} proposed the
first quantum key distribution (QKD) protocol, which was named
BB84 later and on which most the present-day practical
realizations of QKD are based
\cite{bbn,idquantique,magiqtech,toshiba-europe,qinetiq,nec}. In
1991 Eckert proposed a QKD protocol of another type (based on
quantum entanglement) \cite{ekert} which is called E91. There are
also practical realizations of this type of protocols
\cite{zeilinger-ekert}.

Recently the first commercial QKD systems
\cite{idquantique,magiqtech} became available on the market. Some
commercial, military and security institutions are interested in
this new technology. In this connection, the questions about the
concept of the security of QKD protocols and keys generated by
them are crucial.

As with any product the problem of elaborating on some standards
and specifications of QKD systems arises: what kind of
characteristics a producer have to include in the specification.
The necessity of the elaboration of some standards for the
widespread deployment of quantum cryptography has been already
pointed out in \cite{Lo-market}.

In this paper we propose and discuss a list of characteristics for
the specification, which includes a numerical evaluation of the
security of solution and can be considered as a standard for
quantum key distribution solutions. The list is based on the
average time of key generation depending on some parameters.

In the simplest case for the user the list includes three
characteristics: the security degree $\varepsilon$, the length of
keys $m$ and the key refresh rate $R$.

The paper is organized as follows. In Section 2 we remind some
features of QKD. We give the comparison of the computational and
information-theoretic (unconditional) approaches to cryptography,
the notions of QKD protocol and keys security, a classification of
adversary's attacks and some specific features of QKD. In Section
3 we discuss the problem of specification of QKD systems and
propose a list of characteristics of these systems, which can be
taken as a standard and which a producer has to indicate in the
specification.

For a review on quantum cryptography see, e.g.,
\cite{Gisin,VolovichVolovich}.

\section{Features of QKD}

In this section we discuss various properties of QKD which are
relevant to specifications.

\subsection{Computational and information-theoretic approaches}

Two approaches are distinguished in cryptography depending on the
nature of the assumptions about the adversary
\cite{Schneier,Maurer-lect}.

\begin{itemize}

\item \textit{Computational approach} is proposed in \cite{Diffie}
and based on the complexity of solving of some computational
problems (such that, for example, factorization of the whole
numbers or discrete taking the logarithm) and on the assumption
that the adversary's computational power is bounded. However, as
the adversary with the unbounded computational power can solve any
such problem as quickly as he wish and, hence, break the
cryptographic system, computational security is always
conditional. The risk that the security of a system in
computational sense will be broken exists always because of the
progress in the computer engineering (for example, in the
engineering of quantum computers).

\item \textit{Information-theoretic approach} originates in
\cite{Shannon49} and is based on the assumption that the
information of the adversary is bounded. In quantum cryptography
the adversary's information is bounded due to the uncertainty
relations in quantum world. As there are no assumptions on the
adversary's computational power, information-theoretic security is
called \textit{unconditional} and is more desirable.
Theoretically, the adversary has no way to break an
unconditionally-secure cryptographic system, even using infinite
computing power.

\end{itemize}

Most the present-day cryptographic protocols (for example, RSA)
are based on the computational approach, namely, on the lack of
effective algorithms for solution of NP-problems at present.
Besides the weaknesses of the computational approach pointed out
above, the fact that impossibility of the effective solving the
NP-problems isn't proved is considered to be one more weakness of
the present-day cryptosystems. If effective algorithms for solving
NP-problems are found, most of the present-day cryptosystems will
lose their security.

\subsection{Security of pair of keys}\label{SecKeysSecur}

The problem of key distribution is an important problem in
cryptography. Two legal parties, Alice and Bob, want to get a pair
of keys\footnote{In a number of papers it is said about one key,
but we will say about two keys in order to emphasise that formally
there are two different random values and, in general, the Alice's
and Bob's keys don't coincide with each other.} (one key for Alice
and another one for Bob) using communication channels. A
realization of a certain random variable on a finite set $\mathcal
K$, or this random variable itself is regarded as a key. A pair of
keys is called \textit{perfectly secure} if

\begin{enumerate}

\item they are uniformly distributed,

\item they are identical, and

\item a potential adversary (Eve) has no information about them.

\end{enumerate}

Accordingly, the adversary Eve aims, firstly, to get as much
information about the keys as possible and, secondly, to make the
Alice's and Bob's keys different.

It is necessary to evaluate the security of the pair of keys. It
is natural to define the insecurity of the pair of keys as the
distance from the ideal pair of keys which is perfectly secure
\cite{ComposableQKD,ComposablePA}. Since the definition must be
applicable to quantum cryptography, it must be given in terms of
quantum states. Classical state (probability distribution) is a
particular case of quantum state. Let $P_{K_AK_B}$ be a joint
distribution of Alice's key $K_A$ and Bob's key $K_B$. Let $\rho$
be a quantum state which includes both the keys $K_A$ and $K_B$,
and Eve's (in general, quantum) information about these keys. Let
$\rho_{ideal}$ be the state which corresponds to the ideal pair of
keys. Then, the pair of keys $(K_A,K_B)$ is called
$\varepsilon$-\textit{secure} where $\varepsilon\in[0,1]$, if
$$\delta(\rho,\rho_{ideal})\leq1-\varepsilon.$$
The number $\varepsilon$ we will call the \textit{security degree}
of the pair of keys. Here $\delta(\cdot,\cdot)\in[0,1]$ is the
distance measure between two quantum states. So, a pair of keys is
as secure as $\varepsilon$ is closer to 1. A 1-secure pair of keys
is perfectly secure. Note that this security is
information-theoretic (unconditional), because there aren't any
assumptions about Eve's computational power.

See Appendix for the formal definition of the security degree of a
pair of keys. Here we give some important properties of this
definition.

\begin{itemize}

\item This definition of security is \textit{universally
composable} in sense of \cite{ComposabilityQ&C}, which is
important for the modern cryptographic protocols.

\item If the pair of keys is $\varepsilon$-secure, then the
probability $P_{guess}$ that the adversary guesses the keys
(\textit{success probability}) is bounded by (see \cite{PowerQMem}
and \cite{ComposablePA})

$$P_{guess}\leq\frac{1}{|\mathcal K|}+1-\varepsilon$$
where $|\mathcal K|$ is the number of elements in the set of keys
$\mathcal K$. For example, the success probability of the
$\varepsilon$-secure $n$-bit pair of keys is bounded by
$2^{-n}+1-\varepsilon$. If the pair of keys is perfectly secure,
then the success probability is $1/|\mathcal K|$, i.e., the
adversary has no information and can only perform the completely
random guessing from $|\mathcal K|$ elements.

\item The fact that the pair of keys is $\varepsilon$-secure can
be interpreted as that the pair is perfectly secure with the
probability $\varepsilon$.

So, this definition is both useful, because it is universally
composable, and obvious, because it is related to the adversary's
success probability and the probability that the pair is perfectly
secure.


\item If the pair of keys $(K_A^1K_A^2,K_B^1K_B^2)$ which is get
by concatenation of two pairs $(K_A^1,K_B^1)$ and $(K_A^2,K_B^2)$
is $\varepsilon$-secure, then both the pairs $(K_A^1,K_B^1)$ and
$(K_A^2,K_B^2)$ are also $\varepsilon$-secure. The same is hold
for the concatenation of arbitrary number of pairs of keys. So, we
can divide pairs of keys into shorter pairs of keys with the same
degree of security.

\end{itemize}

\subsection{QKD protocol}\label{SecWhatisQKDprotocol}

A practical quantum cryptography system with two legal parties
(Alice and Bob) is a pair of hardware devices (Alice's hardware
device and Bob's one). These devices are connected with each other
by a quantum channel (mostly by optical fiber) and a classical
channel (e.g., Ethernet or optical fiber), and each of the devices
is attached to the corresponding (Alice's or Bob's) computer. It
is clear that for operation of these devices the software is
necessary.

So, in the most general case a quantum cryptography protocol (with
two legal parties) is a pair of programs (algorithms) for a pair
of computers which interact with each other by quantum and
classical channels using special hardware devices. Besides the
commands of a usual programming language, these programs must
contain the additional commands for the hardware devices (lasers,
detectors etc.) management.

Key distribution is one of the problems which can be solved by
quantum cryptography. In QKD, in contrast to other applications of
quantum cryptography, besides the legal parties there is also the
adversary Eve. She also has a computer with an attached hardware
device which allows Eve to eavesdrop the channels and to change
the messages transmitting through them. So, in essence, the
adversary's attack is a program for her computer with a special
hardware. For a more formal discussion see, e.g.,
\cite{RennerEkert,TrushVol}.

The QKD protocols usually include the following steps.

{\renewcommand{\labelenumi}{\theenumi.}
\renewcommand{\theenumi}{\arabic{enumi}}

\begin{enumerate}

\item Photons transmission. Alice transmits to Bob a certain
number of photons through the quantum channel in the states that
she chooses from a certain set. Her choices are unknown to Eve.
Eve can perform different operations on the transmitted photons.
Bob measures these photons in the bases that he also chooses from
a certain set. His choices are also unknown to Eve.

\item Test. Alice and Bob estimate a certain measure of Eve's
interference by analysing the data transmitted and received
through the quantum channel and communicating through the
classical (public) channel. For BB84-type protocols the
\textit{quantum bit error rate} (\textit{QBER}) plays the role of
the measure of Eve's interference. In E91-type protocols the level
of violation of Bell's inequality plays the role of this measure.
Using the estimated value of this measure they estimate Eve's
information about the data. If the estimation of Eve's information
exceeds a certain bound, then Alice and Bob go to step 3. If not,
they go to step 4. This analysis is based on the property of
quantum world: the measurement of the quantum system changes the
state of this system, so it is impossible for Eve to get
information by measuring the transmitted photons without
introducing the noise in them.

\item Decision about the further course of the protocol. The
negative result on step 2 may be caused by both Eve's influence
and statistical fluctuations. Alice and Bob may end the execution
of the protocol, or return to step 1 and run the cycle once again.

\item Classical postprocessing of the quantum data. Alice and Bob
perform the certain classical procedures by communicating through
the classical (public) channel which allow them to correct errors,
to reduce Eve's information and so, to get a pair of keys with the
desirable security degree.

On steps 2 -- 4 Eve taps the classical channel and, may be (see
the next subsection), actively intercept the classical
communication of Alice and Bob.

\end{enumerate}
}

In this way, one gets the quantum state $\rho$ which includes the
Alice's and Bob's keys $K_A$ and $K_B$, and Eve's information
about them, see the previous subsection.

In some cases, steps 3 and 4 can be omitted. For example, in
\cite{Yuen} another approach to quantum cryptography is proposed.
But there is also the same sequence: photon transmission, then
test.

Note the following features of QKD protocols:

\begin{itemize}

\item Probability that Eve guess all Alice's and Bob's choices
during the photon transmission step is negligibly small, but not
zero. In this case, Eve will have full information about the keys.
On the other hand, privacy amplification procedure on step 4 can
reduce Eve's information to arbitrary small amount, but not to
zero. So, a pair of keys generated by QKD system can't be
perfectly secure.

\item It is possible that either Alice or Bob, or both of them
retract the key distribution (see step 3). This is possible even
in case of no eavesdropping, but noisy quantum channel: if there
are too many errors due to the natural noise in the quantum
channel (this can happen with some nonzero probability), Alice and
Bob could think that there is an eavesdropper and retract the key
distribution. In order to reduce this probability (i.e., to reduce
the statistical fluctuations), Alice have to send more photons to
Bob, which has an effect on the time of key generation (see
subsection \ref{SecFCharactQKD}).

\end{itemize}

\subsection{Classification of the adversary's attacks}\label{SecClassAttack}

When we speak about the security degree of the pair of keys which
is generated by the key distribution protocol, we must point out
the class of adversary's attacks relative to which the pair of
keys has the declared security degree. We consider the following
classification.

{\renewcommand{\labelenumi}{\theenumi.}
\renewcommand{\theenumi}{\Roman{enumi}}
\renewcommand{\labelenumii}{(\theenumii)}
\renewcommand{\theenumii}{\roman{enumii}}

\begin{enumerate}

\item By the degree of mastering of quantum technologies by the
adversary.

\begin{enumerate}

\item \textit{Incomplete mastering of quantum technologies}.
Besides the laws of quantum mechanics, there are other
restrictions on the adversary's operations on the photons
transmitted through the quantum channel. For example, the
adversary can perform only individual attacks, or the adversary
can't perform the beam-splitting attack \cite{Gisin}.

\item \textit{Complete mastering of quantum technologies}. During
the photons transmission the adversary can perform with these
photons any operations that are allowed by quantum mechanics.

\end{enumerate}

\item By the authenticity of the classical channel.

\begin{enumerate}

\item \textit{Authentic classical channel}. The adversary can
freely tap the classical channel, but can't change and interrupt
the messages sent by the legal parties, and send other messages.
So, the adversary has read access, but hasn't write access to the
classical channel. In case this assumption, the authenticity of
the channel must be provided by technology.

\item \textit{Unauthentic classical channel}. The adversary can
not only freely tap the classical channel, but also change and
interrupt the messages sent by the legal parties, and send her
messages to Alice and Bob. So, the adversary has read and write
access to the classical channel. In this case, the authenticity of
the channel in the protocol must be provided by mathematics.

Generally speaking, it is more preferable, if the classical
channel isn't assumed to be authentic, but the technological
methods of providing with the authenticity can be in more
effective some cases from the viewpoint of other parameters of the
QKD system (e.g., one can avoid the key degradation problem -- see
subsections \ref{SecKeyDegrad}, \ref{SecFCharactQKD} and
\ref{SecNCharactQKD}).

\end{enumerate}

\item By the adversary's computing power.

\begin{enumerate}

\item \textit{Adversary has limited computing power}.

\item \textit{Adversary has unlimited computing power}.

We should make a remark about the limitation of the computing
power. Assumption about the adversary's computing power can be
applied for using the public-key methods (e.g., digital
signatures) as a mathematical method of authentication of the
classical channel. In this case, the security of the pair of keys
generated by the protocol, generally speaking, isn't
unconditional. But there is an advantage over the public-key
cryptosystems, which is noticed in \cite{WhyQC}. In public-key
cryptosystems, even if the adversary hasn't unlimited computing
power now, in future, when unlimited computing power probably will
be available, she can calculate the secret key using the public
key and so, break the cryptosystem. In case of the use of
public-key methods for authentication in the QKD protocol, if the
adversary hasn't enough computing power now, it is useless to have
unlimited computing power later. So, one can say that, in general,
such a pair of keys is unconditionally-secure against future
attacks.

\end{enumerate}
\end{enumerate}
}

Accordingly, the most general class of attacks is the case when
the adversary has complete mastering of quantum technologies and
unlimited computing power, and the classical channel is
unauthentic.

This classification is rather rough, more precise classifications,
e.g., specifications of adversary's mastering of quantum
technologies (if it is incomplete) and computing power (if it is
limited), are possible. The intermediate authenticity degrees of
the classical channel, e.g., the case when the adversary can send
her messages, but can't change and interrupt other messages (it is
realistic in radio communication), are also possible.

\subsection{Key degradation problem}\label{SecKeyDegrad}

One more significant problem in quantum cryptography is the key
degradation problem, which is considered in \cite{ComposableQKD}.

In case of unauthentic classical channel and unlimited Eve's
computing power, Alice and Bob have to use the unconditional
message authentication codes (MAC), and for that they have to have
a common key, or, as shown in \cite{exactprice}, at least
correlated random variables about which Eve hasn't complete
information. A portion of each of the generated keys must be kept
for the next session, where it will be used as the initial key
(for authentication). However, the obtained pair of keys is not
perfectly secure. So, with every run of the QKD protocol Alice and
Bob obtain less and less secure keys.

Hence, after a number of runs of QKD protocol Alice and Bob need
to obtain a new pair of keys not by QKD protocol. We will call
these keys and source that generates them and deliver to Alice and
Bob \textit{external}. So, in this case, Alice and Bob need to
have an external source of keys.

If the classical channel is authentic, then it's not necessary to
have an external pair of keys. If the channel is unauthentic, but
the Eve's computing power is limited, then Alice and Bob can use
public-key methods for authentication, e.g., digital signatures.
In this case, they need to have an external initial key only at
the beginning for the announcement of the first public key. Then a
portions of public and secret keys is used for the authentication
of the current message, and another portion -- for the
authentication of the announcement of the next public key. In this
case we have no problem of key degradation.

The initial pair of keys can be used not only for the
authentication \cite{Yuen}.

\section{Specifications of QKD systems}

\subsection{Questions to the producers of QKD systems}

At present first commercial QKD systems come into the market
\cite{idquantique,magiqtech}. They provide specifications which
include physical, environmental an some other characteristics of
the QKD systems. Note that for the commercial QKD systems the
length of keys $m$ and the key refresh rate $R$ are indicated
($m=256$ bits, $R=100$ times/second) \cite{idquantique,magiqtech}.
In specifications and descriptions of these systems some important
from practical point of view information is lacking. One asks the
following questions:

\begin{enumerate}

\item How secure can be pair of keys that the user obtain using
these systems?

\item Against which class of attacks are these systems secure?

\item Is the key degradation problem taken into account?

\end{enumerate}

Concerning the security it is claimed that the keys generated by
the commercial QKD systems are absolutely secure. It is not clear
what does it mean. As we have said above, the security degree
$\varepsilon$ of the pair of keys generated by the QKD protocol
can't be equal to 1, i.e., the pair of keys can't be perfectly
secure in this sense. We suggest that such important
characteristic as the security degree of the pair of keys should
be indicated in the specification.

These questions are important since one of the declared advantages
of quantum cryptography over the conventional one is the
availability of rigorous proofs and estimations (see also
discussion in \cite{Yuen}). So, the lack of the rigorous numerical
estimations of the security is a retreat from the original idea of
quantum cryptography. Certainly, any security estimation is
relative: the adversary can perform an attack which is not
concerned directly with the operations on the transmitted photons,
i.e., which isn't taken into account by the mathematical formalism
(the examples of such attacks see, e.g., in \cite{Gisin}).
However, the rigorous numerical security estimations in assumption
that adversary's operations satisfy the declared class of attacks,
in our opinion, are necessary. The estimation of the real security
can be obtained only when numerous various attacks on the
practical QKD systems are carried out. So, we need an army of
"quantum hackers" (see \cite{Lo-market}).

Besides, the following general principle of cryptography is known
\cite{yashchenko,Schneier-vuln}: any statement about the security
of a cryptographic scheme demand the precise specification of
values of all of its parameters, and often even a small deviation
from the established values completely destroys the security of
the system.

\subsection{Maximal measure of Eve's
interference and Success probability in case of no
eavesdropping}\label{SecMaxMEve}

In subsection \ref{SecWhatisQKDprotocol} it was said about the
measure of Eve's interference (QBER for BB84-type protocols and
the level of violation of Bell's inequality for E91-type
protocols). This measure is denoted by $M$.

In \cite{CsiszarKorner78} the notion of secrecy capacity of the
classical broadcast channel was introduced. This is an analogue of
Shannon's channel capacity for the case of the presence of an
eavesdropper: besides the required transmission rate it is
demanded in the definition of secrecy capacity that the
eavesdropper has a negligibly small information. In \cite{Maurer}
these ideas were extended for the case when in addition to the
broadcast channel Alice and Bob can communicate also through the
public channel. The notion of secret key rate was introduced
there.

A quantum channel with classical input (Alice's coding of
classical bits into the quantum states) and classical output
(Bob's and Eve's measurements) can be considered as a classical
broadcast channel. So, in quantum cryptography we can also use the
notion of secret key rate. But, in contrast to the classical
models, in the quantum case the secret key rate $S$ depends on
Eve's activity. Alice and Bob can estimate it by estimating the
measure of Eve's interference $M$, i.e., $S=S(M)$.

And there is a maximal value $M_{max}$ such that $S(M)=0$, if
$M\geq M_{max}$, and $S(M)>0$, if $M<M_{max}$. For example, the
maximal QBER for the BB84 protocol is known to be 11\%
\cite{ShorPreskill}.

This value $M_{max}$ is often used to characterize and compare
different QKD protocols. Larger value of $M_{max}$ for a protocol
means that this protocol is more robust against the natural noise
(i.e., the noise when there is no eavesdropping) in the quantum
channel. If $M_{max}$ is such that due to the natural noise the
value of $M$ estimated by Alice and Bob is more than $M_{max}$
with high probability, then this protocol cannot operate, since
Alice and Bob would think that due to the eavesdropping they
cannot generate a secret key, whereas there is no eavesdropping.

Of course, $M_{max}$ is an important characteristic of a QKD
protocol, but, in our opinion, it has the following drawbacks for
the specification of QKD systems:

\begin{itemize}

\item Secret key rate, as well as secrecy capacity and usually
Shannon's channel capacity, is an asymptotic characteristic: it
guarantees that it is possible to get a pair of keys with security
degree arbitrarily close to one \textit{only for sufficiently
large number of transmitted photons}. But Alice and Bob have only
finite number of transmitted photons on the step of test (see
subsection \ref{SecWhatisQKDprotocol}). If they determine that
this number is not enough to achieve the desired security degree
with the given secret key rate, Alice can transmit more photons to
Bob. But Eve can change her strategy of interception of the
quantum channel and, hence, change the value of secret key rate
during this second transmission. So, the satisfaction of the
condition $M<M_{max}$ does not mean that the distribution of the
pair of keys with the desired security degree is possible;

\item $M_{max}$ is not a universal characteristic of QKD protocol,
since the different measures of Eve's interference are used in the
different protocols;

\item $M_{max}$ is a rather internal characteristic of QKD
protocol. It is of the interest of the engineer who develops the
QKD solution, but not of the engineer who develops further
applications using the QKD solution or of the end-user.

\end{itemize}

$M_{max}$ is not a measure of robustness of the protocol against
Eve's attacks: if Eve wants to break the communication between
Alice and Bob, she can always do it by making $M$ greater than
$M_{max}$. $M_{max}$ is only a measure of robustness of the
protocol against the natural noise. But then we can use the
probability $$\gamma=\Pr[M<M_{max}|\mbox{ no eavesdropping }]$$
instead of $M_{max}$. $\gamma$ is the probability that both Alice
and Bob do not retract the key distribution in case of no
eavesdropping. This parameter is both universal and suitable for
users. We will call $\gamma$ the \textit{success probability in
case of no eavesdropping}.

In fact, $\gamma$ depends on the number $n$ of transmitted
photons: Alice can send more photons in order to decrease the
statistical fluctuations and hence to increase $\gamma$. But we do
not write this dependence (like $\gamma(n)$), because we consider
$\gamma$ as an external parameter, which is set by the user (or it
may be fixed -- see subsection \ref{SecFCharactQKD}), and number
of photons $n$ as an internal parameter of the current operating
of the QKD system, which is not of the interest of the user. So,
the number of photons $n$ depends on $\gamma$. And the computer
program of QKD system determine the required number of photons
$n(\gamma)$ for the given $\gamma$.

\subsection{The simplest specification of QKD parameters}\label{SecSimpSpec}

We propose to use the following three characteristics for the
specification of QKD parameters of the system in the simplest
case:

\begin{itemize}

\item security degree $\varepsilon$,

\item length of keys $m$, and

\item key refresh rate $R$.

\end{itemize}

Here it is assumed that the security degree and the length of keys
in the QKD system are fixed and in this sense this is the simplest
case. If the user can vary $\varepsilon$ and $m$, then
$R=R(m,\varepsilon)$ is a function depending on these parameters.
A pair of keys which is longer or more secure requires more time
for its generation, i.e., the smaller key refresh rate.

But $\varepsilon$ and $m$ are fixed for an individual launch of
the QKD system. So, in all cases these parameters characterise the
individual launch of the QKD system.

Here it is also assumed that there is no key degradation problem
here, i.e., the users don't have to have external keys.

\subsection{Functional engineering characteristics of QKD
systems}\label{SecFCharactQKD}

In this subsection we introduce functional characteristics of the
QKD systems suitable for detailed specifications.

\subsubsection{Average time of key generation}

For the functional description of QKD system we propose to use the
average time of key generation $T$, if the QKD parameters
(security degree, length of keys etc.) are fixed. The average time
$T$ describes the quality of the QKD system. Higher security
requires the longer time of key generation. Note that the time $T$
includes the times required for both photons transmission and
classical computations. That's why we use the time instead of the
number of photons for the description. We suppose that the time
depends on the following parameters:

\begin{enumerate}

\item The desirable length of keys $m$

\item The desirable security degree of keys $\varepsilon$,
$0<\varepsilon\leq 1$

\item The desirable success probability in case of no
eavesdropping (see subsections \ref{SecWhatisQKDprotocol} and
\ref{SecMaxMEve}) $\gamma$, $0<\gamma\leq 1$

\item The length of the initial pair of keys $m_0$, $m_0<m$

\item The security degree of the initial pair of keys.
$\varepsilon_0$, $0<\varepsilon_0\leq 1$.

\end{enumerate}

So, the average time $T$ is a function
$T=T(m,\varepsilon,\gamma,m_0,\varepsilon_0)$. The average time
$T$ increases as $m$, $\varepsilon$ or $\gamma$ increase, or $m_0$
or $\varepsilon_0$ decrease.
$T(m,\varepsilon,\gamma,m_0,\varepsilon_0)=\infty$ is interpreted
as an impossibility of key generation with the given parameters.
Here it is assumed that the distance of the QKD system functioning
is fixed.

\subsubsection{Key refresh rate and key generation rate}

The average time $T$ describes the QKD system in details, but it
depends on too many arguments. It is necessary to introduce
functions which describes the QKD system not so detailed, but have
less parameters. One of these characteristics is the \textit{key
generation rate}. In order to define the key generation rate
properly we must analyse what information is needed for user.

In the following we fix $\gamma$, say, $\gamma=0,99$, and do not
consider the dependence of the functional characteristics below on
$\gamma$. Furthermore, for simplicity we at first consider the
case when $m_0=0$ (no key degradation). So, the average time $T$
depends only on two arguments: the desirable length of keys $m$
and the desirable security degree of the pair of keys
$\varepsilon$. We will write $T(m,\varepsilon)$.

The user is interested in the pair of values $(m,\varepsilon)$,
i.e., he want to generate a pair of keys (only once or
continuously) with the length $m$ bits and the security degree
$\varepsilon$. And he want to know the time $T(m,\varepsilon)$ (is
measured, e.g., in seconds) during which he can generate it. Or,
equivalently, he want to know the \textit{key refresh rate}
$$R(m,\varepsilon)=\frac{1}{T(m,\varepsilon)}$$
(is measured in times/second), which is more common in
specifications of key distribution systems. We must give the
definition of the key generation rate so that the user knowing the
key generation rate and the desirable parameters $(m,\varepsilon)$
could find (may be approximately) the key refresh rate. It is
natural to define the key generation rate as
$$\tilde V(m,\varepsilon)=\frac{m}{T(m,\varepsilon)}=mR(m,\varepsilon)$$ (is measured in
bits/second). But we want to eliminate the length $m$ from the
arguments of the key generation rate, because it is natural to
define the key generation rate which depends on the security
degree, but does not dependent on the length. We define the key
generation rate as
$$V(\varepsilon)=\lim_{m\to\infty}\frac{m}{T(m,\varepsilon)}$$
(it is assumed that the limit, may be infinite, exists).

Explain the introduced definition. Let $(m,\varepsilon)$ is the
desired pair of parameters.
$$V(\varepsilon)\approx\frac{m_\infty}{T(m_\infty,\varepsilon)}$$
where $m_\infty$ is a large number. Let $m_\infty=mn$ where $n$ is
some natural number. We divide the pair of keys with the length
$m_\infty$ into $n$ pairs of keys with the length $m$. The
security of each of these shorter pairs is also $\varepsilon$ (see
the properties of the security degree at the end of subsection
\ref{SecKeysSecur}). So, $n$ pairs of keys with the length $m$ and
the security degree $\varepsilon$ are generated during the time
$T(m_\infty,\varepsilon)\approx m_\infty/V(\varepsilon)$. The key
refresh rate is $R(m,\varepsilon)=n/T(m_\infty,\varepsilon)\approx
V(\varepsilon)/m$. Thus, the user knowing $(m,\varepsilon)$ and
$V(\varepsilon)$ can calculate the key refresh rate by the formula

\begin{equation*}
R(m,\varepsilon)\approx\frac{V(\varepsilon)}{m}.\end{equation*}

It is possible that in concrete QKD systems there are faster ways
for generating keys with the parameters $(m,\varepsilon)$ than
generating much longer keys with the same security degree. But the
value $V(\varepsilon)/m$ gives the guaranteed key refresh rate.

If $V(\varepsilon)=\infty$, then arbitrarily large key refresh
rates are achievable by the proper (large enough) choice of
$m_\infty$.

Now consider the general case where
$T=T(m,\varepsilon,\gamma,m_0,\varepsilon_0)$ ($\gamma$ is fixed
as before). Now Alice and Bob must have an external pair of keys
in order to generate a pair of longer keys. So, the key refresh
rate
$$R(m,\varepsilon,m_0)=\frac{1}{T(m,\varepsilon,\gamma,m_0,1)}$$
has an additional parameter: the length of initial (external) keys
$m_0$, i.e., the length of the perfectly secure keys that Alice
and Bob must have before the QKD session in order to generate the
keys with the length $m$ and the security degree $\varepsilon$.
Smaller $m_0$ is more desirable, but it can decrease the key
generation rate. For example, some security degrees becomes
unavailable (i.e., the key refresh rate falls to zero) when the
length of the external pair of keys becomes too small. Or more
rounds in the authentication protocol \cite{Auth-interact}, which
require additional time, are needed in order to generate keys with
the same security degree, but having the external pair of keys
with a shorter length.

So, $R(m,\varepsilon,m_0)=r$ times/second means that Alice and Bob
using the QKD system can refresh $r$ times per second their keys
with the length $m$ and security degree $\varepsilon$, and before
each refreshing they must have for that at the average $m_0$ bits
of the perfectly secure external keys (if the external keys are
not perfectly secure, then they must be longer than $m_0$).

Now in order to define the key generation rate we should take
$m\to\infty$ and $m_0\to\infty$ so that $\frac{m_0}{m}=D=const$.
The amount $D$ we will call \textit{external key consumption
rate}. Thus, in this case the key generation rate depends on two
parameters: the security degree $\varepsilon$ and the external key
consumption rate $D$. Since $m_0=\lfloor Dm\rfloor$, where
$\lfloor x\rfloor$ denotes the floor of the real number $x$, i.e.,
the nearest to $x$ integer from below, we define the key
generation rate as

$$V(\varepsilon,D)=\lim_{m\to\infty}\frac{m}{T(m,\varepsilon,\gamma,\lfloor
Dm\rfloor,1)}.$$

Knowing $m$, $m_0$ and $V(\varepsilon,\frac{m_0}{m})$ one can
calculate $R(m,\varepsilon,m_0)$ by the formula

\begin{equation}\label{EqRviaV}
R(m,\varepsilon,m_0)\approx\frac{V(\varepsilon,\frac{m_0}{m})}{m}
\end{equation}

Of course, it makes no sense to decrease the security degree to
$0$, or to increase the external key consumption rate to $1$ or
greater (in the first case the optimal way for Alice and Bob is to
generate two keys independently, in the latter case the optimal
way is to use the external pair of keys for the direct purpose
instead of generation a pair with a shorter length). So, the
domain of the function $V(\varepsilon,D)$ is $0<\varepsilon\leq
1$, $0\leq D<1$.

By implication, $V$ is a continuous function on its domain, a
non-increasing function of $\varepsilon$ and a non-decreasing
function of $D$.

\subsubsection{Upper bound of security degrees}

One more important functional characteristic of QKD system is the
upper bound of the security degrees which can be achieved with the
given external key consumption rate. It can't decrease as external
key consumption rate $D$ increases. We define this function
$\varepsilon_{max}(D)$ of $D$, $0<\varepsilon\leq 1$, by the
following formula:

$$\varepsilon_{max}(D)=\min\{\varepsilon|V(\varepsilon,D)=0\}.$$

By implication, $\varepsilon_{max}$ is a continuous function on
its domain and a non-decreasing function of $D$.

Since there is the security degree among the arguments of the
functions $T$ and $V$, it is necessary to point out the class of
attacks (see subsection \ref{SecClassAttack}) relative to which
the keys have the declared security.

In view of the key degradation problem (see subsection
\ref{SecKeyDegrad}) we will distinguish the systems with
\textit{one-time} and \textit{permanent external key consumption}.
Formally, we will say that the system needs maximum one-time
external key consumption (no key degradation problem), if
$V(\varepsilon,D)=const$ when $\varepsilon$ is fixed (hence,
$\varepsilon_{max}(D)=const$). Otherwise we will say that the
system needs permanent external key consumption.

\subsection{Numeric engineering and end-user characteristics of
QKD systems}\label{SecNCharactQKD}

Thus, for engineering description of the QKD system we have
proposed the functions
$T=T(m,\varepsilon,\gamma,m_0,\varepsilon_0)$, $V(\varepsilon,D)$,
$\varepsilon_{max}(D)$.

It is worthwhile to simplify these functional characteristics to a
set of numerical characteristics of QKD systems which may be
useful both for engineers and end-users. So, there is a problem to
choose a set of numbers which good describes the functions.

\subsubsection{No key degradation case}

At first, we consider the simple case of one-time external key
consumption rate, i.e.,
$V(\varepsilon,D)=V(\varepsilon,0)=V(\varepsilon)$ and
$\varepsilon_{max}(D)=\mathrm{MAXS}=const$.

It is clear that we are interested in the generation of keys with
at most achievable security degrees. We are not interested in the
behaviour of the function $V(\varepsilon)$ in the area where
$\varepsilon$ is close to zero. So, we must choose some numerical
characteristics of the function $V(\varepsilon)$ which concerns
with the interesting area.

First, we are interested in the upper bound $\mathrm{MAXS}$ of the
achievable security degrees. By  continuity of the function $V$,
$V(\mathrm{MAXS})=0$, so we can generate keys only at rates
smaller than $\mathrm{MAXS}$. But there is a difference how fast
increases the rate as the security degree decreases from
$\mathrm{MAXS}$. In order to characterise this we approximate the
function $V(\varepsilon)$ by its tangent in the point
$\mathrm{MAXS}$ and introduce the \textit{marginal increment of
key generation rate} ($\mathrm{MIR}$)
$$\mathrm{MIR}=-\left.\frac{dV(\varepsilon)}{d\varepsilon}\right|
_{\varepsilon=\mathrm{MAXS}}$$ where the derivative is left-sided
(since $V(\varepsilon)$ hits zero in $\mathrm{MAXS}$, this point
may be salient). Since $V(\varepsilon)$ is a non-increasing
function, $dV(\varepsilon)/d\varepsilon\leq0$ and
$\mathrm{MIR}\geq0$. So, key generation rate with the desirable
security degree of the pair of keys $\varepsilon$ is approximately
calculated by
$$V(\varepsilon)\approx\mathrm{MIR}(\mathrm{MAXS}-\varepsilon).$$
Vice versa, the security degree of the pair of keys generated at a
given rate $V$ is calculated by
$$\varepsilon(V)\approx\mathrm{MAXS}-\frac{V}{\mathrm{MIR}}.$$

So, in this simple (but practical) case, we need only two numbers
in order to approximately characterise a QKD system: the
\textit{marginal security degree} $\mathrm{MAXS}$,
$0<\mathrm{MAXS}\leq1$ and the marginal increment of key
generation rate $\mathrm{MIR}$, $0\leq\mathrm{MIR}<\infty$. If a
QKD system has greater $\mathrm{MAXS}$ and $\mathrm{MIR}$ than
another one, then the first QKD system is better because it allows
to generate more secure keys at higher rates.

\subsubsection{The general case}

Now we consider the general case. We are interested in the key
generation with the maximal security degree and the minimal
external key consumption rate, i.e., in the area where
$\varepsilon$ is close to $1$ and $D$ is close to $0$. But the
difficulty is that $\varepsilon(0)$ can be far from $1$ and this
is an unacceptable variant. So, the user have to find compromise
values of $\varepsilon$ and $D$.

The value that characterise the quality of the QKD system is the
minimal achievable distance of the curve $\varepsilon_{max}(D)$,
$0\leq D<1$, to the point $(\varepsilon=1,D=0)$:

$$\mathrm{DIST}=\min_{0\leq
D<1}\sqrt{a(\varepsilon_{max}(D)-1)^2+bD^2}$$ where $a>0$ and
$b>0$, $a+b=1$, are some fixed coefficients. For example, one can
take $a=b=0.5$. But it may be useful to set $a$ and $b$ so that
$a>b$, because the security degree and the external key
consumption rate are not equivalent amounts. For example, $D=0,1$
(i.e., on each 10 bits of the new keys one have to spend 1 bit of
the external keys) may be acceptable, but the security degree
$\varepsilon=1-0,1=0,9$ may be too small. If $a>b$, one pay larger
penalty for the distance $\varepsilon$ from 1 than for the
distance $D$ from 0. Optimal values of $D$ and $\varepsilon$ with
respect to this distance we denote by $D^*$ and
$\varepsilon^*=\varepsilon_{max}(D^*)$. These values can also be
used as a characteristic of the QKD system: these are the
parameters at which key generation is optimal.

But $V(\varepsilon^*, D^*)=0$ by definition of the function
$\varepsilon_{max}(D)$ and continuity of the function
$V(\varepsilon, D)$. As in the case $D=0$, we have to introduce a
characteristic showing how fast the key generation rate increases
when $\varepsilon$ decreases from $\varepsilon^*$ and $D$ remains
constant. We define marginal increment (MIR) of key generation
rate as

$$\mathrm{MIR}=-\left.\frac{\partial V(\varepsilon,D)}{\partial\varepsilon}\right|
_{(\varepsilon^*,D^*)}$$ where the derivative is left-sided.

We are also interested in the ends of the function
$\varepsilon_{max}(D)$. Consider the right end. There are two
possibilities for: either $\varepsilon_{max}(0)>0$ or
$\varepsilon_{max}(0)=0$. In the latter case, define
$$D_{min}=\inf\{D|\varepsilon_{max}(D)>0\}.$$

Of course, the first case is more preferable than the second one.
In the first case, it's possible to generate pairs of keys with
some security degree without external key consumption. In the
second case, the external key consumption rate can't be smaller
than $D_{min}$ even if we want to generate a pair of keys with
very small security degree. We define the quantity
$$\mathrm{SOC}=\begin{cases}-D_{min},&\text{if $D_{min}>0$}\\
\varepsilon_{max}(0),&\text{if $D_{min}=0$}\end{cases},$$ which we
will call the \textit{security degree of the pair of keys without
the external key consumption}. Negative $\mathrm{SWE}$ corresponds
to the second case where the external key consumption rate can't
be smaller than some value ($-\mathrm{SWE}$).

Similarly analyse the right end of the function
$\varepsilon_{max}(D)$, i.e.,
$$\varepsilon_{max}(1)\stackrel{def}{=}\lim_{D\to1}\varepsilon_{max}(D).$$

There are also two possibilities: $\varepsilon_{max}(1)=1$ or
$\varepsilon_{max}(1)<1$. In the first case, define
$$D_{max}=\min\{D|\varepsilon_{max}(1)=1\}.$$

The first case is more preferable than the second one. In the
first case, it is possible to generate keys with the security
arbitrarily close to perfect with external key consumption rate
less than the amount $D_{max}\leq 1$. In the second case, the
security degree can't be larger as $\varepsilon_{max}(1)<1$ even
if the external key consumption rate is very close to 1. We define
the quantity
$$\mathrm{GMC}=\begin{cases}-(1-\varepsilon_{max}(1)),
&\text{if $\varepsilon_{max}(1)<1$}\\
1-D_{max},&\text{if $\varepsilon_{max}(1)=1$}\end{cases},$$ which
we will call the \textit{gain at the maximal external key
consumption}.

Thus, we obtain six characteristics of the QKD system with the
external key consumption: $DIST$, $\varepsilon^*$, $D^*$,
$\mathrm{MIR}$, $\mathrm{SOC}$ and $\mathrm{GMC}$.

Approximately the key generation rate in a point $(\varepsilon,
D)$, $\varepsilon\leq\varepsilon_{max}(D)$ is given by

$$V(\varepsilon,D)\approx\mathrm{MIR}(\varepsilon_{max}(D)-\varepsilon)$$

It is assumed that the user generates the keys with the parameters
near the optimal point $(\varepsilon^*,D^*)$, so
$\varepsilon_{max}(D)\approx\varepsilon_{max}(D^*)=\varepsilon^*$.
And the user knowing the above numeric characteristics can
approximately (rather rough) calculate the key generation rate in
a point $(\varepsilon, D)$, $\varepsilon\leq\varepsilon_{max}(D)$,
by the formula

\begin{equation}\label{EqVviaMIR&EpsOpt}
V(\varepsilon,D)\approx\mathrm{MIR}(\varepsilon^*-\varepsilon)
\end{equation}
Vice versa, the security degree of pair of keys generated at a
given rate $V$ and external key consumption rate $D$ is calculated
by
$$\varepsilon(V,D)\approx\varepsilon^*-\frac{V}{\mathrm{MIR}}.$$
$D^*$ is an approximate value of the external key consumption
rate, if the user generates the keys in a point near the optimum.
$\mathrm{SOC}$ and $\mathrm{GMC}$ don't participate in these
approximations, but they characterise the potential abilities of a
QKD system. And $\mathrm{DIST}$ is an index of quality of a
system.

Consider the simple case of no external key consumption from the
point of view of the general case. It was said before that
$V(\varepsilon,D)=const$, when $\varepsilon$ is fixed, and
$\varepsilon_{max}(D)=const=\mathrm{MAXS}$. Evidently, $D^*=0$,
$\varepsilon^*=\mathrm{MAXS}=\mathrm{SOC}$ and
$\mathrm{DIST}=1-\mathrm{MAXS}$.
$\mathrm{GMC}=-(1-\mathrm{MAXS})$, if $\mathrm{MAXS}<1$, and
$\mathrm{GMC}=1$, if $\mathrm{MAXS}=1$. The quantity
$\mathrm{MIR}$ coincides with the same quantity that we defined
for the simple case. Thus, one can use these characteristics for
both the general and simple cases.

\subsection{The list of characteristics for the
specification}\label{SecListCharactQKD}

Of course, besides these numeric characteristics the user must
know about the assumptions about the adversary and the distance
within which these characteristics are valid. Finally, we propose
the following list of qualitative and quantitative characteristics
which can be included in the specification:

\begin{enumerate}

\item The assumed degree of the adversary's mastering of quantum
technologies: incomplete/complete

\item Method of providing with the authenticity of the classical
channel: technological/mathematical

\item The assumed adversary's computing power: limited/unlimited

\item Distance from the ideal $\mathrm{DIST}$ (variation interval
is $[0,1)$, dimensionless value)

\item The optimal security degree $\varepsilon^*$ (variation
interval is $(0,1]$, dimensionless value)

\item The optimal external key consumption rate $D^*$ (variation
interval is $[0,1)$, dimensionless value)

\item Marginal increment of the key generation rate $\mathrm{MIR}$
(variation interval is $(0,\infty)$ bit/sec)

\item Security degree of the pair of keys without the external key
consumption $\mathrm{SOC}$ (variation interval is $(-1,1]$,
dimensionless value)

\item Gain at the maximal external key consumption $\mathrm{GMC}$
(variation interval is $(-1,1]$, dimensionless value)

\item The distance within which these characteristics are valid
(km).

\end{enumerate}

Larger value of each of the numeric characteristics (except (iv))
is preferable. In the first three (qualitative) characteristics
the second value is preferable.

It is assumed that the producer of the QKD system has to give to
the engineer the functions $t$, $V$, $\varepsilon_{max}$
(analytical formulas or graphics) and characteristics 1 -- 10. To
the end-user the producer has to give characteristics 1 -- 10.

The present-day commercial quantum cryptography solutions have the
encryption systems (AES and 3DES) attached to the QKD systems. The
security of these encryption protocols when the keys are perfectly
secure is a problem of conventional cryptography, but the above
(or similar) characteristics about the security of keys and key
generation must be given.

In subsection \ref{SecSimpSpec} we have introduced three
characteristics for the simplest case: security degree
$\varepsilon$, length of keys $m$ and key refresh rate $R$. Length
of keys $m$ drops out since $m$ is not a constant any more: in the
general case the user can choose any $m$. Security degree
$\varepsilon$ is also not a constant any more, but some
information about the values that $\varepsilon$ can have is given
in characteristics (iv) and (v) (in the case of no key degradation
problem, these characteristics are equal). $R(m,\varepsilon)$ as a
function of $m$ and $\varepsilon$, which specifies the user, can
be calculated by formulas (\ref{EqRviaV}) and
(\ref{EqVviaMIR&EpsOpt}).

For the end-user ten characteristics may be too many and it's
necessary to reduce the number of characteristics. Firstly, some
of the above characteristics may be equal for all or for a very
wide class of the QKD systems and will be eliminated. Secondly,
some of these characteristics may be for engineers rather then for
end-users. In our opinion, characteristics (i)-(iv), (vii) and (x)
(i.e., three qualitative and three quantitative characteristics)
are most important for the user.

\section{Acknowledgements}

This work was partially supported by the Russian Foundation of
Basic Research (project 05-01-00884), the grant of the President
of the Russian Federation (project NSh-6705.2006) and the program
"Modern problems of theoretical mathematics" of the Mathematical
Sciences department of the Russian Academy of Sciences.

\appendix

\section*{Appendix. Definition of the security degree of pair of keys}
\addcontentsline{toc}{bibl}{\normalsize\bfseries{Appendix.
Definition of the security degree of pair of keys}}

\begin{definition}[see \cite{ComposableQKD,ComposablePA}]\label{defkeysecur}Let $\mathcal K$ be a finite or a countable set,
$K_A,K_B$ be a pair of random variables (keys) on $\mathcal K$
with the joint distribution $P_{K_A,K_B}$. Let, further,
$\mathcal{H}_{AB},\mathcal{H}_E$ be Hilbert spaces,
$\dim\mathcal{H}_{AB}=|\mathcal K|^2$,
$\{|k_A,k_B\rangle\}_{k_A,k_B\in\mathcal K}$ be an orthonormal
base of $\mathcal{H}_{AB}$. The pair of keys $(K_A,K_B)$ is called
$\varepsilon$-\textit{secure} relative to the joint (with the
adversary) quantum state

$$\rho=\sum_{k_A,k_B\in\mathcal
K}P_{K_AK_B}(k_A,k_B)|k_A,k_B\rangle\langle
k_A,k_B|\otimes\rho_{k_A,k_B}^E\in\mathcal{S(H}_{AB}\otimes\mathcal{H}_E)$$where
$$\rho_{k_A,k_B}^E\in\mathcal{S(H}_E),k_A,k_B\in\mathcal K,$$
if
$$\delta(\rho,\rho_{ideal})\leq 1-\varepsilon$$
where
$$\rho_{ideal}=\left(\sum_{k\in\mathcal K}\frac{1}{|\mathcal
K|}|k,k\rangle\langle
k,k|\right)\otimes\left(\sum_{k_A,k_B\in\mathcal
K}P_{K_AK_B}(k_A,k_B)\rho_{k_A,k_B}^E\right).$$
\end{definition}

Here $\delta(\cdot,\cdot)$ is the distance between two quantum
states. For arbitrary $\sigma,\eta\in\mathcal{S(H)}$ where
$\mathcal H$ is a Hilbert space,

$$\delta(\sigma,\eta)=\|\sigma-\eta\|_1\stackrel{def}{=}
\sum_{\lambda\in\spec(\sigma-\eta)}|\lambda|.$$

Let $\mathcal X$ be a finite set. The \textit{variational
distance} between two probability distributions (classical states)
$P$ and $Q$ on this set

$$\delta(P,Q)=\frac{1}{2}\sum_{x\in\mathcal X}|P(x)-Q(x)|$$
is the classical analogue and a particular case of the above
distance between quantum states.

For the distance $\delta(\cdot,\cdot)$ the following properties
are satisfied. $\mathcal{H,H}'$ are arbitrary Hilbert spaces and
$\sigma,\eta\in\mathcal{S(H)}$, $\sigma',\eta'\in\mathcal{S(H}')$
are arbitrary states.

\begin{enumerate}

\item
\begin{equation*}
\delta(\sigma\otimes\sigma',\eta\otimes\eta')\leq\delta(\sigma,\eta)+
\delta(\sigma',\eta')\end{equation*}with equality if
$\sigma'=\eta'$.

\item For arbitrary function (quantum operation) $\mathcal E$ on
$\mathcal{S(H)}$
$$\delta(\mathcal E(\sigma),\mathcal E(\eta))\leq\delta(\sigma,\eta)$$
As a particular case, if $\mathcal H=\mathcal H_1\otimes\mathcal
H_2$, $\sigma=\sigma_1\otimes\sigma_2$,
$\eta=\eta_1\otimes\eta_2$, $\sigma_1,\eta_1\in\mathcal H_1$,
$\sigma_2,\eta_2\in\mathcal H_2$ and $\mathcal
E(\sigma_1\otimes\sigma_2)=\sigma_1$, $\mathcal
E(\eta_1\otimes\eta_2)=\eta_1$, then
$$\delta(\sigma_1,\eta_1)\leq\delta(\sigma_1\otimes\sigma_2,\eta_1\otimes\eta_2).$$
It implies that we can divide the pairs of keys into shorter pairs
of keys with the same degree of security (see subsection
\ref{SecKeysSecur}).

\item Consider the probability distributions $P$ and $Q$ of the
outcomes when the same measurement to $\sigma$ and $\eta$,
respectively, is applied. Then
$$\delta(P,Q)\leq\delta(\sigma,\eta).$$

\end{enumerate}

\end{document}